\documentclass[reprint,amsmath,amssymb,aps,usenatbib,useAMS,nofootinbib,floatfix]{revtex4-1}

\usepackage{graphicx}
\usepackage{dcolumn}
\usepackage{bm}
\usepackage{color}
\usepackage{lipsum}
\usepackage{multirow}
\usepackage{centernot}
\usepackage[dvipsnames,table]{xcolor}
\usepackage{hyperref}
\hypersetup{urlcolor=RoyalBlue,citecolor=ForestGreen,linkcolor=RedViolet,colorlinks=true}
\usepackage[normalem]{ulem}
\usepackage{aas_macros}
\usepackage{gensymb}
\usepackage{multirow}
\usepackage{float}
\usepackage[left=1.9cm,right=1.9cm,top=2cm,bottom=2.5cm]{geometry}

\newcommand{\eg}{\textit{e.g.}}

\definecolor{background}{RGB}{255,255,255} 

\begin{document}
\pagecolor{background}

\title{Anisotropic Counts-in-Cells in Redshift Space:\\ A New Route to Cosmological Constraints from Galaxy Surveys}

\author{Pawe\l{} Drozda} \email{(pdrozda, hellwing, bilicki)@cft.edu.pl}
\author{Wojciech A. Hellwing}
\author{Maciej Bilicki}
\affiliation{Center for Theoretical Physics, Polish Academy of Sciences, Al. Lotników 32/46,02-668 Warsaw, Poland}


\date{\today}

\begin{abstract}
We introduce a novel extension of the volume-averaged correlation function (VACF) framework by replacing the traditional spherical smoothing kernels with anisotropic, ellipsoidal windows. This generalized approach enables the study of shape-dependent clustering statistics and captures directional information encoded in large-scale structure, particularly in redshift space where galaxy distribution is distorted by peculiar velocities. We define and compute ellipsoidal VACFs \(\bar{\xi}_J(r_{\parallel}, r_{\perp})\) and the corresponding reduced cumulants \(s_J(r_{\parallel}, r_{\perp})\), allowing for joint sensitivity to both scale and anisotropy across arbitrary statistical order \(J\). Using a suite of COLA N-body simulations spanning a grid of cosmologies with varying \(\Omega_M\) and \(\sigma_8\), we analyze the behavior of ellipsoidal VACFs and cumulants in both real and redshift space. We find that the shape of the smoothing kernel that maximizes the clustering signal depends strongly on the redshift-space distortion regime: spherical in real space, prolate in the Fingers-of-God-dominated regime, and oblate in the Kaiser squashing-dominated regime. While the standard VACF amplitude is mainly sensitive to \(\sigma_8\), the shape-dependence of redshift-space skewness shows a coherent response to the combined growth parameter \(f\sigma_8\), with a typical sensitivity at the 1–3\(\sigma\) level between neighboring models. Our results demonstrate that ellipsoidal VACFs offer a computationally efficient and information-rich generalization of counts-in-cells analysis, with promising applications to galaxy survey data, halo catalogs, and cosmological tests of gravity beyond \(\Lambda\)CDM.
\end{abstract}

\maketitle

\section{Introduction}\label{sec.intro}
The large-scale structure (LSS) of the Universe originates from the gravitational amplification of primordial density perturbations, modulated by baryonic processes in an expanding cosmological background. The current standard model of cosmology, \(\Lambda\)CDM, successfully describes a broad range of observations across cosmic time. Nonetheless, persistent tensions—such as those concerning the Hubble constant and the growth rate of structure—motivate the investigation of extensions to the standard model, including modifications to gravity and the nature of dark energy.

The two-point correlation function (2PCF) and its Fourier counterpart, the power spectrum, remain central tools in the statistical analysis of LSS. While powerful, these second-order statistics are inherently limited in their ability to capture non-Gaussian features and the full nonlinear complexity of cosmic structure formation. To address this, various higher-order statistics—such as three-point correlation function and bispectrum , marked correlation functions, and hierarchical cumulants—have been developed, albeit at considerable computational cost \citep[\eg][]{Scoccimarro2000,bernardeau2002large,skibba2006the}.

The use of correlation functions in cosmology was pioneered by Peebles in the 1970s and 1980s \citep{peebles1980large}. Building on the fair-sample hypothesis, volume-averaged correlation functions (VACFs) were later introduced as the moments of the counts-in-cells (CiC) distribution, offering a more tractable alternative to high-order n-point functions. This framework was significantly advanced in the 1990s, mostly through the work of Juszkiewicz, Bernardeau, Colombi, and Gaztañaga, who demonstrated that VACFs and their reduced combinations could be used to test the gravitational instability paradigm and to constrain cosmological parameters \citep{juszkiewicz1993skewness,ber92,colombi1994hierarchy,gaztanaga1994high}. In subsequent decades, CiC-based statistics have been widely employed to explore a range of alternative cosmological scenarios, including modified gravity models \citep[\eg][]{Hellwing2013, Hellwing2017, Alam2021,Cataneo2022}.

In practice, VACFs are computed as the central moments of tracer counts within smoothing volumes of varying size—typically spherical. This traditional implementation, while computationally efficient and conceptually clear, suffers from a key limitation: the use of spherical kernels imposes isotropy and suppresses directional information. As a result, it fails to capture the anisotropic features of the Cosmic Web, such as the elongated morphology of filaments or the flattened structure of walls \citep[][]{Hivon1995, Codis2016, Repp2020, Drozda2025, Gould2025}. Furthermore, in spectroscopic surveys, peculiar velocities introduce anisotropies along the line-of-sight (LOS), giving rise to redshift-space distortions (RSD). These distortions encode additional cosmological information—especially about the growth rate of structure—but their directional signature is diluted by isotropic averaging.

This issue is especially pressing in the era of precision cosmology. Galaxy spectroscopic surveys provide three-dimensional maps of the cosmic density field with high fidelity, particularly along the LOS. Past and current datasets such as 2dFGRS, SDSS, or GAMA, along with next-generation surveys including Euclid, DESI, and 4MOST, offer unprecedented volumes and sizes of spectroscopic samples \citep[\eg][]{Weinberg2013, DESICollaboration2016, Taylor2023, Euclid2025, DiValentino2025}. These data present a unique opportunity to perform high-precision CiC analyses that fully exploit their anisotropic information content. However, doing so requires extending the traditional isotropic framework to account for LOS-dependent effects introduced by RSD.

While RSD signals have been extensively utilized in two-point statistics to constrain the growth rate and galaxy bias, their incorporation into higher-order CiC statistics has remained underdeveloped. A natural and long-overdue extension of the CiC formalism is to introduce anisotropic smoothing volumes that are sensitive to redshift-space effects.
In this work, we address this gap by generalizing the VACF methodology to include ellipsoidal smoothing kernels with independently controlled axes parallel and perpendicular to the LOS. This extension allows us to explicitly probe shape-dependent clustering, capture anisotropies induced by cosmic velocities, and gain access to information otherwise suppressed by spherical averaging.

We apply this new estimator to a suite of N-body simulations based on PiCOLA (\textit{Parallel COmoving Lagrangian Accelerator}) dynamics\citep{Howlett2015}, exploring a grid of cosmological models with varying matter density \(\Omega_M\) and fluctuation amplitude \(\sigma_8\). By analyzing anisotropic VACFs and their associated cumulants in both real and redshift space, we uncover distinctive morphological patterns and RSD signatures that trace the growth of structure. Our results show that ellipsoidal VACFs provide a powerful and flexible tool for extracting anisotropic clustering information from spectroscopic surveys.

This paper is organized as follows. In Section~\ref{sec.calc}, we define the ellipsoidal VACF estimator and describe its numerical implementation. Section~\ref{sec.data} presents the simulation data and construction of redshift-space catalogs. In Section~\ref{sec.res}, we present our results for real- and redshift-space VACFs and cumulants across a range of cosmological models. Finally, in Section~\ref{sec.concl}, we summarize our findings and discuss their implications for future analyses of large-scale structure.

\section{Shape-dependent density field statistics}\label{sec.calc}

The starting point of our analysis is the matter density contrast field, defined as
\begin{equation}
    \delta(\vec{x}) = \frac{\rho(\vec{x})}{\bar{\rho}} - 1,
\end{equation}
where \(\rho(\vec{x})\) is the local matter density at position \(\vec{x}\), and \(\bar{\rho}\) is the mean cosmic density. 

To probe the statistical properties of the density field beyond the two-point level, we consider the volume-averaged correlation function (VACF) of order \(J\), defined as
\begin{equation}
    \bar{\xi}_J(R) \equiv \langle \bar{\delta}^J \rangle,
\end{equation}
where \(\bar{\delta}(\vec{x})\) denotes the density contrast field smoothed over a finite volume using a three-dimensional window function \(W_{3D}\),
\begin{equation}
    \bar{\delta}(\vec{x}) = \int d^3 x' \, \delta(\vec{x}') \, W_{3D}(\vec{x} - \vec{x}').
\end{equation}

In the standard formulation, \(W_{3D}\) is taken to be a spherically symmetric top-hat kernel of radius \(R\) \citep{peebles1980large, bernardeau2002large}. Under the assumption that the smoothed local density can be approximated by counting tracers (e.g. halos or galaxies) within spheres of radius \(R\), the VACF can be estimated as the central moments of the CiC distribution.
As stated earlier,
in this work we generalize the VACF formalism by introducing an anisotropic smoothing kernel with ellipsoidal geometry.
We define the ellipsoidal window function as
\begin{equation}
W_{3D}(\vec{x}; r_{\parallel}, r_{\perp}) =
\begin{cases}
  \dfrac{3}{4\pi r_{\parallel} r_{\perp}^2}, & \text{if } \dfrac{x_{\parallel}^2}{r_{\parallel}^2} + \dfrac{x_{\perp,1}^2 + x_{\perp,2}^2}{r_{\perp}^2} \leq 1 \\
  0, & \text{otherwise,}
\end{cases}
\label{eq.W3D}
\end{equation}
where \(r_{\parallel}\) and \(r_{\perp}\) denote the semi-axes of the ellipsoid aligned parallel and perpendicular to the LOS, respectively. The position vector \(\vec{x}\) is decomposed as \(\vec{x} = (x_{\parallel}, x_{\perp,1}, x_{\perp,2})\), where we assume symmetry in the plane perpendicular to the LOS, i.e. \(r_{\perp,1} = r_{\perp,2} = r_{\perp}\). This assumption preserves azimuthal symmetry around the LOS and ensures that rotational orientation within the perpendicular plane does not affect the clustering measurement.

The resulting generalized VACF is thus a function of both ellipsoid axes:
\begin{equation}
    \bar{\xi}_J(r_{\parallel}, r_{\perp}) \equiv \langle \bar{\delta}^J \rangle \quad \text{with } W_{3D} \text{ as in Eq. \eqref{eq.W3D}}.
\end{equation}

In practice, we estimate \(\bar{\xi}_J(r_{\parallel}, r_{\perp})\) by computing the central moments of the distribution of tracer counts within ellipsoidal volumes defined by \((r_{\parallel}, r_{\perp})\). From these raw moments, we further derive the connected moments and apply Poisson noise corrections following the formalism of \citet{gaztanaga1994high}.

\begin{figure*}
\centering
\includegraphics[width=\textwidth]{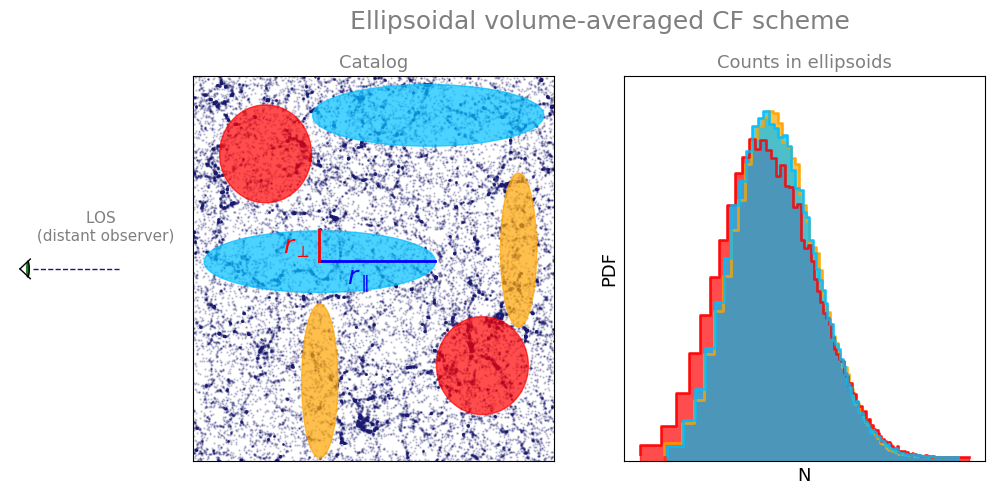}
\caption{Illustration of the ellipsoidal counts-in-cells procedure used to compute \(\bar{\xi}_J(r_{\parallel}, r_{\perp})\). For each pair of kernel axes \((r_{\parallel}, r_{\perp})\), ellipsoidal shells are tiled across the simulation volume. The resulting count distribution is used to compute connected and Poisson-corrected central moments.}
\label{fig.info}
\end{figure*}

A schematic illustration of the procedure is shown in Fig.\ \ref{fig.info}. For each pair \((r_{\parallel}, r_{\perp})\), we compute counts-in-ellipsoids across the simulation volume, derive the moments, and assemble the ellipsoidal VACFs.\footnote{The corresponding code is publicly available at \href{https://github.com/Pawel-96/Avcorr}{\texttt{github.com/Pawel-96/Avcorr}}.}

In addition to the raw and connected moments, we compute reduced cumulants—also known as hierarchical amplitudes—defined as
\begin{equation}
    S_J(r_{\parallel}, r_{\perp}) = \frac{\bar{\xi}_J(r_{\parallel}, r_{\perp})}{\bar{\xi}_2^{J-1}(r_{\parallel}, r_{\perp})}, \quad \text{for } J \geq 3.
\end{equation}
These normalized quantities remove the dominant scale dependence of \(\bar{\xi}_2\) and highlight deviations from Gaussianity and linear growth. In particular, the reduced third-order cumulant or skewness, \(S_3 = \bar{\xi}_3 / \bar{\xi}_2^2\), is frequently used to test the predictions of gravitational instability and probe higher-order mode coupling,\citep{juszkiewicz1993skewness,juszkiewicz2013skewness}. In the weakly nonlinear regime, \(S_J\) are predicted to be nearly scale-independent
\cite{bouchet1992weakly}
and are sensitive to both cosmological parameters and the shape of the initial power spectrum 
\citep{bernardeau1994skewness,gaztanaga1998the}.

\section{Data}\label{sec.data}
This work is based on a new suite of gravity-only N-body simulations generated using the COLA (COmoving Lagrangian Acceleration) method \citep{tassev2013solving}, in particular we used the MG-COLA variant \cite{winther2017cola}.
These simulations were specifically constructed to explore the impact of cosmological parameters on anisotropic clustering and higher-order statistics. They represent the first application of this COLA ensemble, which is optimized to strike a balance between computational efficiency and sufficient resolution for analyzing the matter density field.

The fiducial cosmology is based on the Planck 2018 results \citep{planck2020planck}, corresponding to the TT,TE,EE+lowE+lensing+BAO solution from Table 2 of \citet{planck2020planck}.
The adopted cosmological parameters are:
\begin{itemize}
    \item Hubble parameter: \( h = 0.6766 \),
    \item Baryon density: \( \Omega_b = 0.049 \),
    \item Total matter density: \( \Omega_m = 0.3111 \),
    \item Dark energy density: \( \Omega_\Lambda = 0.6889 \),
    \item Power spectrum normalization: \( \sigma_8 = 0.8102 \),
    \item Spectral index: \( n_s = 0.9665 \).
\end{itemize}

Each simulation evolves \(1024^3\) dark matter particles in a periodic box of size \(L_{\rm box} = 500\,h^{-1}\mathrm{Mpc}\), starting the time integration at redshift \(z = 19\) and integrated using 52 global time steps. Forces are computed on a regular mesh with \(1536^3\) cells, yielding a spatial resolution of approximately 325 kpc/\(h\). This level of resolution is well suited for studies of the continuous matter density field, though it is not sufficient for resolving halo substructure or detailed halo assembly histories. A dedicated analysis of halo and galaxy clustering using higher-resolution simulations will be presented in a follow-up work.

The full suite consists of 9 distinct background cosmologies, spanning \(\pm10\%\) variations in \(\Omega_m\) and \(\sigma_8\) around the fiducial model (see Table \ref{tab.models}). For each cosmology, we generate 5 independent realizations using different initial condition seeds, resulting in a total of 45 simulations. This design ensures robust statistical power for measuring redshift-space clustering and cosmological trends.

We analyze three redshift snapshots: \(z = \{0.1, 0.2, 0.5\}\), using dark matter pseudo-particles as tracers. All statistics are computed in the distant observer approximation \citep{kaiser1987clustering,scoccimarro2004redshift,taruya2010baryon},
without applying light-cone projection or relativistic corrections.

To facilitate interpretation and comparison with growth-based observables, we compute the derived combination
\begin{equation}
    f\sigma_8 = f(\Omega_m)\,\sigma_8 \approx \Omega_m^{0.55} \sigma_8,
\end{equation}
where \(f \equiv d\ln D / d\ln a\) is the linear growth rate. This parameter is frequently used in redshift-space distortion analyses as a robust measure of structure growth \citep{guzzo2008a,blake2011the,beutler2012the}.

\begin{table}[t]
\caption{Cosmological models used in this study, with specified values of \(\sigma_8\), \(\Omega_M\), and the derived \(f\sigma_8\). The fiducial model is denoted as \texttt{Cen}.}
\begingroup
\setlength{\tabcolsep}{5pt}
\renewcommand{\arraystretch}{1.3}
\begin{tabular}{c c c c}
\hline
Model & \(\sigma_8\) & \(\Omega_M\) & \(f\sigma_8\) \\
\hline\hline
\texttt{Ls8COM}  & 0.7292 & 0.3111 & 0.3837 \\
\texttt{Ls8HOM}  & 0.7292 & 0.3422 & 0.4043 \\
\texttt{Cs8LOM}  & 0.8102 & 0.2800 & 0.4023 \\
\texttt{Cen}     & 0.8102 & 0.3111 & 0.4263 \\
\texttt{Cs8HOM}  & 0.8102 & 0.3422 & 0.4492 \\
\texttt{Hs8LOM}  & 0.8912 & 0.2800 & 0.4425 \\
\texttt{Hs8COM}  & 0.8912 & 0.3111 & 0.4689 \\
\texttt{Hs8HOM}  & 0.8912 & 0.3422 & 0.4941 \\
\hline
\end{tabular}
\endgroup
\label{tab.models}
\end{table}

In addition to real-space catalogs, we construct redshift-space versions of each snapshot using the distant observer approximation. For each tracer particle, the redshift-space position is computed as
\begin{equation}
    \vec{x}_{\rm zspace} = \vec{x} + \frac{1 + z_{\rm snap}}{H(z_{\rm snap})} \, v_{\parallel} \, \hat{e}_{\parallel},
\end{equation}
where \(v_{\parallel}\) is the peculiar velocity component along the LOS and \(\hat{e}_{\parallel}\) is the LOS unit vector. This transformation introduces anisotropies from both coherent infall (Kaiser effect) and small-scale dispersion (Finger-of-God effect) \citep{hamilton1998linear}.

To average over orientation-dependent effects, we generate three redshift-space realizations for each snapshot by aligning the LOS direction with each Cartesian axis in turn. All redshift-space measurements are averaged over these three direction-projections and over 5 initial phase retaliations, following standard practice \citep{jennings2011simulations,hellwing2020the}.

\section{Anisotropic counts-in-cells}\label{sec.res}

\begin{figure}[t]
\centering
\includegraphics[width=\columnwidth]{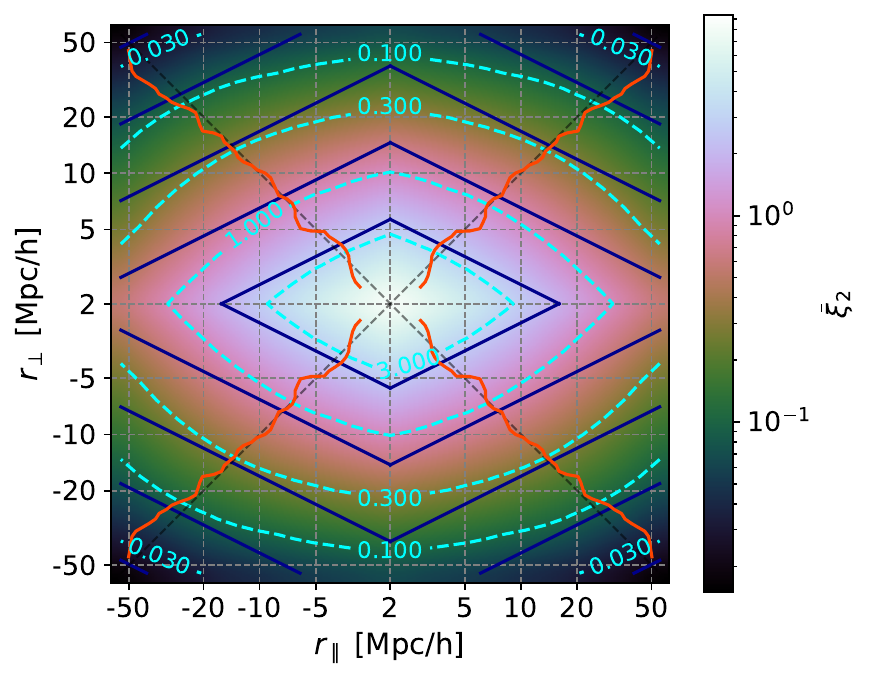}
\caption{Second-order ellipsoidal VACF \(\bar{\xi}_2(r_{\parallel}, r_{\perp})\) in real space at redshift \(z = 0.1\). Dashed cyan lines denote iso-correlation contours; dark blue diamond-shaped curves indicate iso-volume contours (\(4\pi/3 \cdot r_{\parallel} r_{\perp}^2 = \mathrm{const}\)); orange-red lines trace the ellipsoidal shapes \([r_{\parallel}, r_{\perp}]\) that maximize \(\bar{\xi}_2\) at fixed volume.}
\label{fig.W2_DM_REAL}
\end{figure}

We begin our analysis by examining the geometric and physical properties of ellipsoidal volume-averaged correlation functions (VACFs). As a starting point, we consider the second-order VACF, \(\bar{\xi}_2(r_{\parallel}, r_{\perp})\), measured in the real space for the fiducial cosmological model using dark matter particles. The result is shown in Figure~\ref{fig.W2_DM_REAL}.

To aid interpretation, we highlight two important geometric structures in the \((r_{\parallel}, r_{\perp})\) plane:

\begin{itemize}
  \item \textbf{Iso-volume contours} connect ellipsoidal kernels of constant volume, defined by \(V = (4\pi/3) \cdot r_{\parallel} r_{\perp}^2\). These contours allow one to compare the clustering signal across different shapes while holding the enclosed volume fixed.
  
  \item \textbf{Iso-correlation contours} follow lines of constant VACF value, \(\bar{\xi}_2(r_{\parallel}, r_{\perp}) = \text{const}\). Their geometry reveals how clustering strength depends jointly on the kernel size and anisotropy.
\end{itemize}

In an isotropic and homogeneous matter distribution, these two families of contours would be expected to align. Thus, any misalignment between them signals the presence of structure in the cosmic density field. Indeed, Figure~\ref{fig.W2_DM_REAL} shows that the iso-correlation lines (dashed cyan) systematically deviate from the iso-volume contours (dark blue), reflecting the anisotropic nature of the Cosmic Web. As expected, the amplitude of \(\bar{\xi}_2\) decreases with increasing scale. Along the diagonal \(r_{\parallel} = r_{\perp}\), corresponding to spherical kernels, the VACF reduces to its standard isotropic form which we studied in
\cite{drozda2025skewness}. However, for fixed volume, the clustering signal is consistently maximized for spherical kernels, as indicated by the orange-red curves. This behavior is in line with the statistical isotropy of over- and under-dense regions in real space: spherical kernels best isolate coherent structures, while elongated kernels tend to span across different environments, suppressing the variance in counts and hence reducing \(\bar{\xi}_2\).

We also observe a symmetry in the VACF response to ellipsoidal shape: for fixed volume \(V\), kernels with axis ratio \(\eta = r_{\parallel}/r_{\perp}\) yield the same VACF as those with inverse ratio \(1/\eta\),
\begin{equation}
    \bar{\xi}_2(r_{\parallel}/r_{\perp} = \eta)\big|_V = \bar{\xi}_2(r_{\parallel}/r_{\perp} = 1/\eta)\big|_V.
    \label{eq.inverseratio}
\end{equation}
This symmetry arises from the isotropy of real space, where no direction is privileged. Consequently, the VACF depends only on the kernel’s shape and volume, not on its orientation with respect to the observer.

At higher orders \(J > 2\), similar trends persist, though the statistical noise increases due to the growing sensitivity of higher-order moments to sample variance. Moreover, the location of maximum \(\bar{\xi}_J\) at fixed volume begins to drift away from the diagonal as non-Gaussian features become more prominent.

\begin{figure}[t]
\centering
\includegraphics[width=\columnwidth]{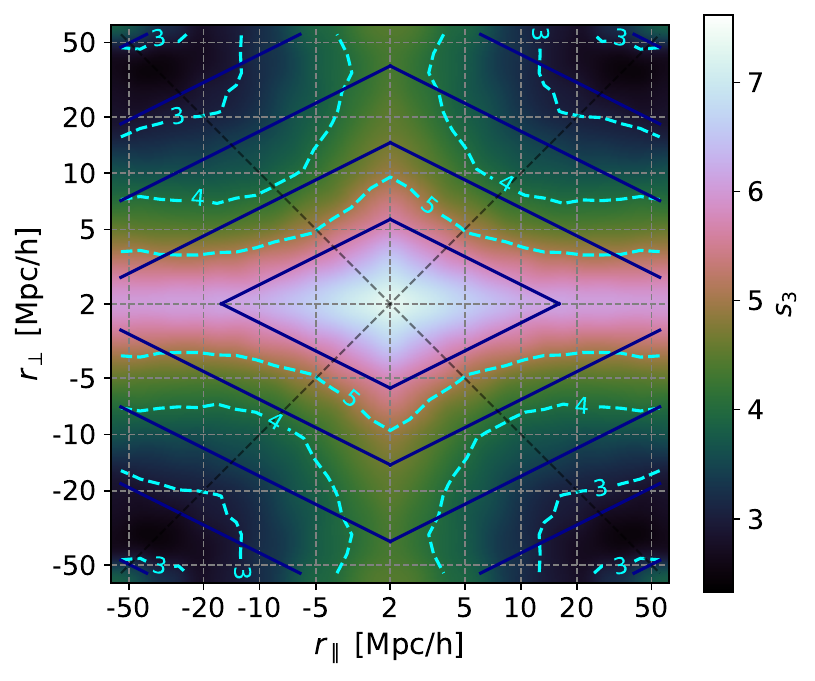}
\caption{Skewness \(S_3\) of the density field in real space as a function of ellipsoidal kernel shape at redshift \(z = 0.1\).}
\label{fig.S3_DM_REAL}
\end{figure}

To gain deeper insight, we now turn to the skewness statistic. Figure~\ref{fig.S3_DM_REAL} shows the third-order reduced cumulant, \(S_3\), for the same dark matter sample in real space.

In perturbation theory (PT), the reduced cumulants \(S_J\) are proportionality factors relating \(\bar{\xi}_J\) to powers of \(\bar{\xi}_2\), i.e., \(S_J \propto \bar{\xi}_J / \bar{\xi}_2^{J-1}\) \citep{fry1993biasing,bernardeau2002large}. In the linear regime, these ratios are scale-independent and determined by the gravitational growth dynamics
\citep{bernardeau2002large}. Deviations from this behavior indicate nonlinear evolution or the breakdown of PT assumptions.

Several key features emerge in Figure~\ref{fig.S3_DM_REAL}. First, \(S_3\) is the largest at small scales, where nonlinear gravitational clustering enhances higher-order correlations. Second, strong variations appear along the axes \(r_{\parallel} = \mathrm{const}\) and \(r_{\perp} = \mathrm{const}\), resulting in a cross-shaped structure. This behavior stems from the asymmetric response of \(\bar{\xi}_3\) and \(\bar{\xi}_2\) to changes in the ellipsoid’s geometry.

In particular, we find that for fixed \(r_{\perp}\) and increasing
\(r_{\parallel} \ll r_{\perp}\), the third-order moment \(\bar{\xi}_3\) declines more rapidly than \(\bar{\xi}_2^2\), leading to a local enhancement in \(S_3\). A similar effect occurs for fixed \(r_{\parallel}\) and small \(r_{\perp} \ll r_{\parallel}\). This results in skewness enhancements along the vertical and horizontal axes. Interestingly, the effect is stronger along the horizontal axis (\(r_{\perp}\)), which reflects the geometry of the smoothing kernel: the ellipsoidal volume depends quadratically on \(r_{\perp}\), making the statistic more sensitive to its variation.

\begin{figure}[t]
\centering
\includegraphics[width=\columnwidth]{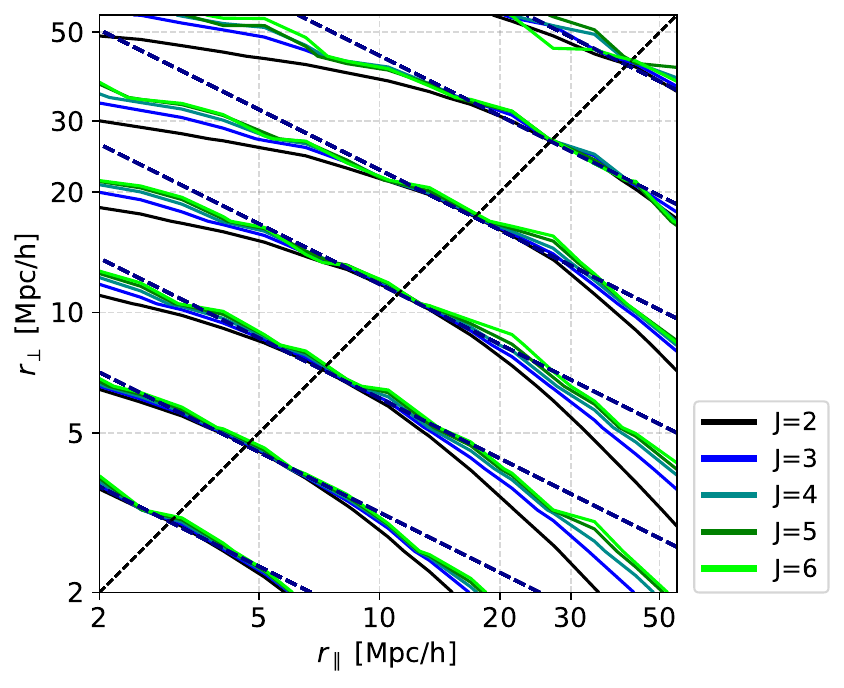}
\caption{Iso-value contours of \(\bar{\xi}_J(r_{\parallel}, r_{\perp})\) for orders \(J = 2, ..., 6\), color-coded as indicated. Dashed dark blue lines mark iso-volume contours passing through \(r_{\parallel} = r_{\perp}\). Only one quadrant is shown for clarity.}
\label{fig.WJ_contours_DM}
\end{figure}

This cross-like morphology in \(S_3\) reflects a broader trend observed across higher-order VACFs. In Figure~\ref{fig.WJ_contours_DM}, we plot the iso-value contours of \(\bar{\xi}_J(r_{\parallel}, r_{\perp})\) for several orders \(J = 2\) through \(6\).

As \(J\) increases, the iso-\(\bar{\xi}_J\) contours converge toward the iso-volume lines (dashed blue diagonals in the Figure). This convergence explains why all reduced cumulants \(S_J\) exhibit similar cross-like features: the higher the order, the more closely \(\bar{\xi}_J\) depends only on the total volume, with diminishing sensitivity to ellipsoid shape. For \(J \to \infty\), one might expect \(\bar{\xi}_J(r_{\parallel}, r_{\perp})\) to become purely volume-dependent.

This behavior is somewhat counterintuitive. One might expect higher-order statistics—sensitive to rare events and extreme structures—to exhibit greater anisotropy. Yet our results suggest that the geometric dependence of \(\bar{\xi}_J\) becomes increasingly isotropic with order. A possible explanation may lie in the slower evolution of higher-order nonlinearities, or in their reduced sensitivity to small-scale anisotropies, but further theoretical investigation is needed to clarify this effect.

\subsection{Redshift-space effects} \label{sec.rsd}

We now turn to the redshift space, the natural domain of galaxy spectroscopic surveys and the primary motivation for our anisotropic CiC estimator. In this context, the parameterization \([r_{\parallel}, r_{\perp}]\) acquires particular relevance, as it allows us to separate the clustering signal along and across the LOS.

To facilitate comparison across different ellipsoidal volumes, we define an effective radius:
\begin{equation}
    r_{\mathrm{eff}} \equiv \left( r_{\parallel} r_{\perp}^2 \right)^{1/3},
\end{equation}
which corresponds to the radius of a sphere with the same volume as the ellipsoid.

\begin{figure}[t]
\centering
\includegraphics[width=\columnwidth]{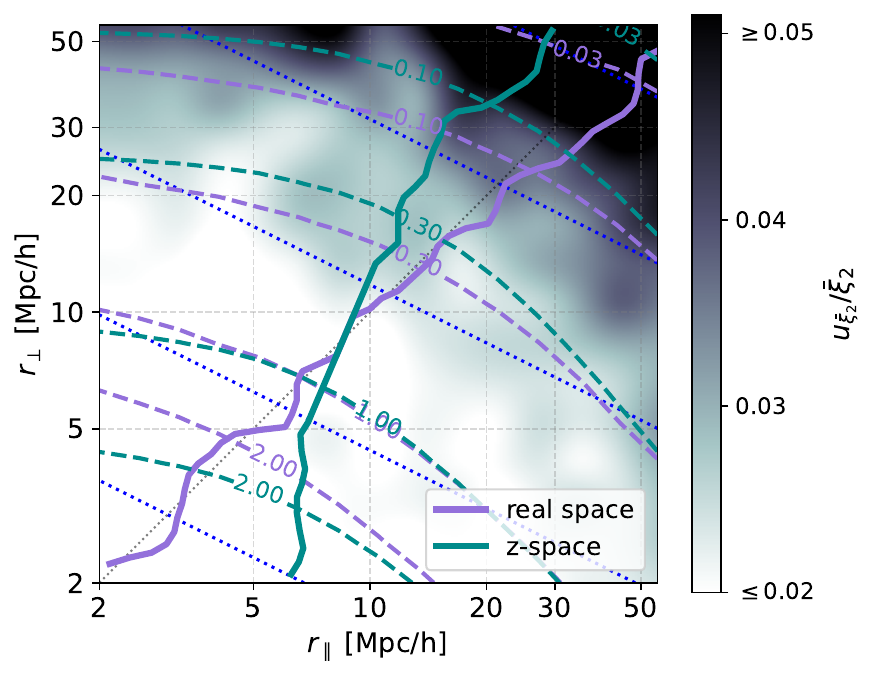}
\caption{Contour plot of dark matter VACF \(\bar{\xi}_2(r_{\parallel}, r_{\perp})\) in real (dashed violet) and redshift space (dashed cyan) at \(z = 0.1\). Solid lines indicate kernel shapes that maximize \(\bar{\xi}_2\) at constant ellipsoidal volume. Dotted blue lines denote iso-volume contours. The background color map shows the average relative uncertainty, \(u_{\bar{\xi}_2}/\bar{\xi}_2\), combining real and redshift space errors.}
\label{fig.W2_contours}
\end{figure}

Figure~\ref{fig.W2_contours} shows iso-correlation contours of the second-order VACF, \(\bar{\xi}_2(r_{\parallel}, r_{\perp})\), in both real and redshift space at redshift \(z = 0.1\). The solid lines indicate the ellipsoidal shapes that maximize \(\bar{\xi}_2\) at fixed volume.

Several key trends emerge. In real space, the VACF is maximized by spherical kernels (\(r_{\parallel} = r_{\perp}\)), consistent with the statistical isotropy of density fluctuations. In redshift space, however, this symmetry is broken by redshift-space distortions (RSD), leading to a scale-dependent shift in the optimal kernel shape.

At small scales (\(r_{\mathrm{eff}} \lesssim 10~h^{-1}\mathrm{Mpc}\)), the maximal signal is achieved for prolate ellipsoids elongated along the LOS (\(r_{\parallel} > r_{\perp}\)). This reflects the dominance of the Fingers-of-God (FoG) effect 
\cite{jackson1972a,dodelson2003modern}, where random virial motions elongate structures along the LOS, smearing the density field. At larger scales, the Kaiser squashing effect
\cite{scoccimarro2004redshift,percival2009testing}
dominates: coherent infall velocities compress structures along the LOS, and the VACF is maximized for oblate ellipsoids with \(r_{\parallel} < r_{\perp}\). The transition between these regimes occurs near \(r_{\mathrm{eff}} \approx 12~h^{-1}\mathrm{Mpc}\), consistent with previous studies of redshift-space clustering \citep[e.g.,][]{jackson1972a,kaiser1987clustering,shi2016mapping}.

To build physical intuition, it is helpful to recall the standard configuration-space prediction for RSD in linear theory. In the Kaiser model\citep{kaiser1987clustering}, the redshift-space two-point correlation function \(\xi^s(r, \mu)\) depends on the separation \(r\) and the cosine \(\mu\) of the angle between the pair vector and the LOS:
\begin{equation}
    \xi^s(r, \mu) = \sum_{\ell=0,2,4} \xi_\ell(r) L_\ell(\mu),
\end{equation}
where \(L_\ell(\mu)\) are Legendre polynomials and the multipoles \(\xi_\ell(r)\) are given by:
\begin{align}
\xi_0(r) &= \left(1 + \frac{2\beta}{3} + \frac{\beta^2}{5}\right) \xi(r), \\
\xi_2(r) &= \left(\frac{4\beta}{3} + \frac{4\beta^2}{7}\right) \left[\xi(r) - \bar{\xi}(r)\right], \\
\xi_4(r) &= \frac{8\beta^2}{35} \left[\xi(r) + \frac{5}{2}\bar{\xi}(r) - \frac{7}{2}\bar{\bar{\xi}}(r)\right],
\end{align}
with \(\beta = f/b\) the linear redshift-space distortion parameter. Here, \(\xi(r)\) is the real-space two-point correlation function, while \(\bar{\xi}(r)\) and \(\bar{\bar{\xi}}(r)\) are its first and second volume-averaged integrals:
\begin{align}
\bar{\xi}(r) &= \frac{3}{r^3} \int_0^r \xi(s) s^2 \, ds, \\
\bar{\bar{\xi}}(r) &= \frac{5}{r^5} \int_0^r \xi(s) s^4 \, ds.
\end{align}

Our ellipsoidal VACF estimator probes the same anisotropic effects encoded in \(\xi^s(r, \mu)\), but from a complementary angle: by systematically varying the shape of the counting kernel and averaging over all directions, we recover the imprint of RSD without requiring explicit pair counts or angular binning.

Notably, the redshift-space VACF does not preserve the symmetry described in Eq.~\ref{eq.inverseratio}, as the LOS direction introduces a preferred axis. This breaks the equivalence between ellipsoids with axis ratios \(\eta\) and \(1/\eta\), leading to asymmetric behavior in the \([r_{\parallel}, r_{\perp}]\) plane.

We also observe that, at small scales, the redshift-to-real-space ratio of \(\bar{\xi}_2\) is suppressed (\(< 1\)), while at larger scales it becomes enhanced (\(> 1\)). This agrees with well-known results for the monopole component of the two-point function and its volume-averaged variants \citep[e.g.,][]{peacock1994reconstructing,scoccimarro2004redshift,taruya2010baryon}.

Furthermore, the ellipsoidal formalism allows us to identify the shapes that yield the most extreme differences between real and redshift space. At small volumes, oblate ellipsoids (\(r_{\parallel} < r_{\perp}\)) provide the lowest signal in redshift space, amplifying the FoG suppression relative to real space. At large volumes, those same oblate shapes yield the strongest enhancement, aligning with the Kaiser effect. Interestingly, the maximal redshift-to-real-space ratio is not achieved at the most oblate shapes, suggesting that the mapping \(\bar{\xi}_2(r_{\parallel}/r_{\perp})\) is asymmetric and has a well-defined peak but no true minimum within the physical domain \(r_{\parallel} > 0\).

These findings demonstrate that ellipsoidal VACFs encode rich anisotropic information beyond that available from spherical statistics. The additional degrees of freedom—smoothing scale and shape—enable a more nuanced analysis of large-scale structure morphology and redshift-space distortions.

\subsection{Growth rate dependence}

In the previous sections, we examined the behavior of ellipsoidal volume-averaged correlation functions and their cumulants in both real and redshift space. We now turn to investigating how the anisotropic VACFs respond to changes in the growth rate of cosmic structure, as encoded by variations in the cosmological parameters \(\Omega_M\) and \(\sigma_8\).

\begin{figure}[t]
\centering
\includegraphics[width=\columnwidth]{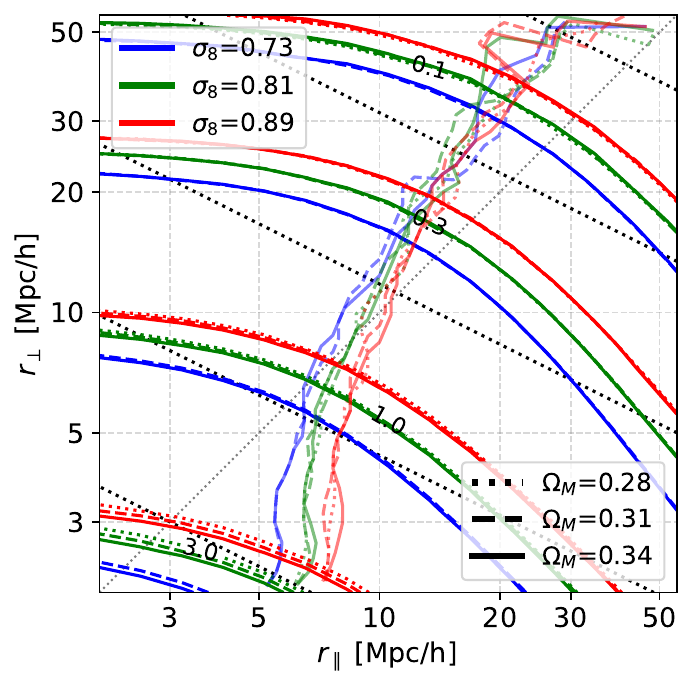}
\caption{Redshift-space \(\bar{\xi}_2(r_{\parallel},r_{\perp})\) at \(z = 0.1\). Colored iso-value contours correspond to different \(\sigma_8\) values (blue, green, red), while line styles (dotted, dashed, solid) denote different values of \(\Omega_M\). Semi-transparent lines trace the ellipsoidal shapes that maximize \(\bar{\xi}_2\) at fixed volume for each model.}
\label{fig.W2_comparefs8}
\end{figure}

Figure~\ref{fig.W2_comparefs8} shows iso-correlation contours and lines of maximal \(\bar{\xi}_2(r_{\parallel}, r_{\perp})\) at constant effective volume for a set of cosmological models sampled at redshift \(z = 0.1\), all computed in redshift space.

As expected, variations in \(\sigma_8\) have a stronger effect on the amplitude of the correlation function than changes in \(\Omega_M\). This reflects the role of \(\sigma_8\) in normalizing the matter power spectrum, to which \(\bar{\xi}_2\) is directly related. In contrast, \(\Omega_M\) influences the shape of the power spectrum and the growth rate \(f\), resulting in more subtle changes to \(\bar{\xi}_2\).

At scales smaller than the transition threshold \(r_{\mathrm{eff}} \lesssim 10~\mathrm{Mpc}/h\), increasing \(\Omega_M\) leads to a suppression of the clustering signal. This is consistent with stronger redshift-space distortions—specifically the FoG effect—expected for models with faster growth rates. On larger scales, where the Kaiser effect dominates, the trend reverses slightly: models with larger \(\Omega_M\) exhibit slightly enhanced \(\bar{\xi}_2\), although the effect is weaker.

Interestingly, the location of the maximum \(\bar{\xi}_2\) at fixed volume shifts with \(\sigma_8\), especially on small scales. Higher \(\sigma_8\) models tend to peak for more prolate ellipsoids (\(r_{\parallel} > r_{\perp}\)), consistent with stronger FoG elongations along the LOS. Conversely, for volumes beyond \(r_{\mathrm{eff}} \sim 10\ \mathrm{Mpc}/h\), models with lower \(\sigma_8\) exhibit maxima at more oblate shapes. This counterintuitive behavior arises because in high-\(\sigma_8\) models, strong small-scale FoG suppression persists even when averaged over larger volumes.

At higher redshifts (not shown here), we observe a systematic shift of the maxima toward more oblate ellipsoids. This trend reflects a reduction in the FoG effect and an enhancement of the Kaiser effect at earlier cosmic times, when coherent infall motions were more prominent. This evolution is in agreement with theoretical expectations (e.g., \cite{guzzo2008a}).

While the amplitude of \(\bar{\xi}_2\) is clearly sensitive to \(\sigma_8\), its dependence on the combined parameter \(f\sigma_8 = \Omega_M^{0.55} \sigma_8\) is non-monotonic. For example, the model Cs8LOM (\(f\sigma_8 = 0.4023\)) yields a stronger \(\bar{\xi}_2\) than Ls8HOM (\(f\sigma_8 = 0.4043\)), while in other cases, such as between Hs8COM and Hs8HOM, the ranking is reversed. This degeneracy makes \(\bar{\xi}_2\) a poor discriminator of \(f\sigma_8\).

\begin{figure}[t]
\centering
\includegraphics[width=\columnwidth]{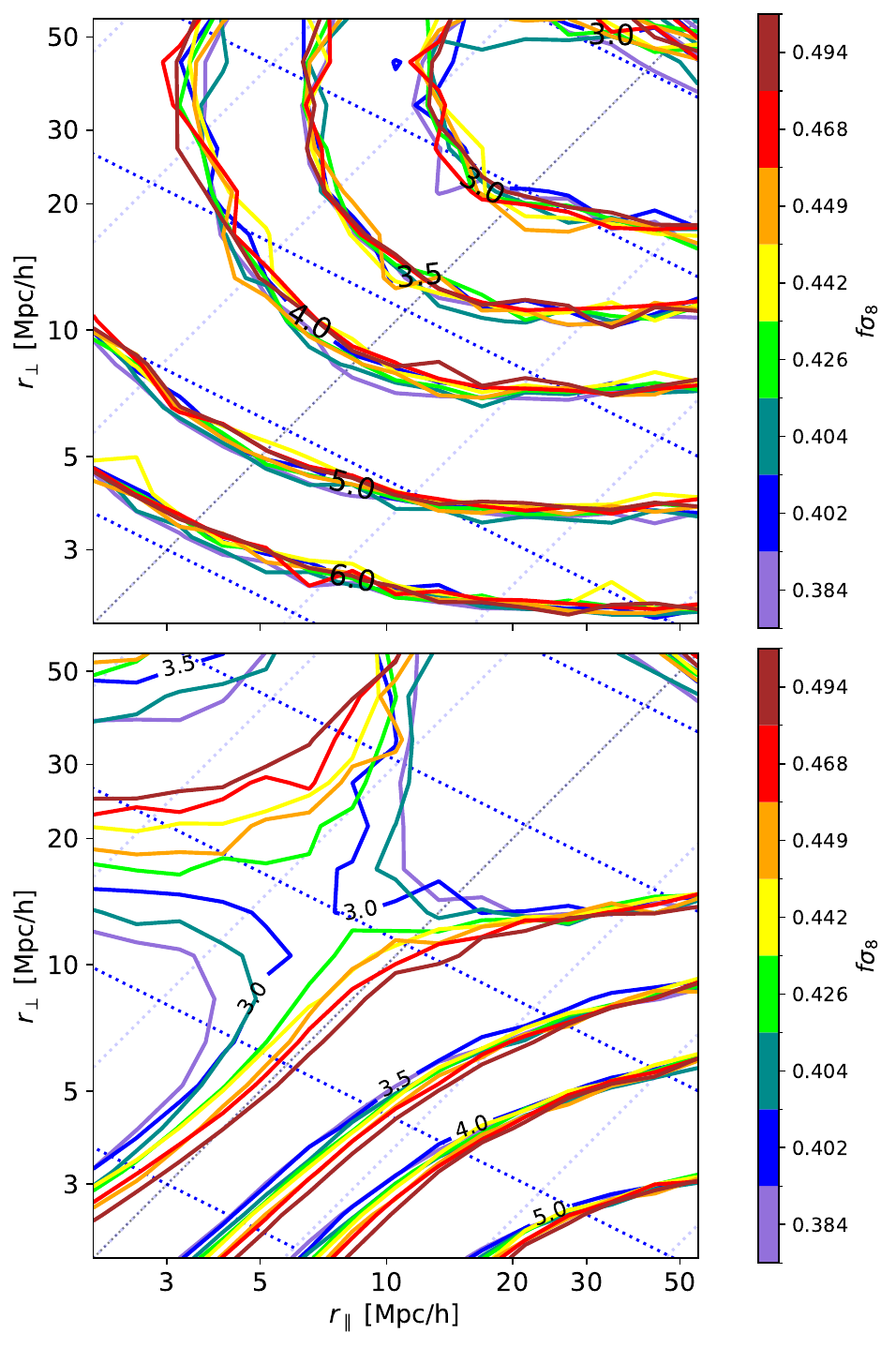}
\caption{Iso-contours of reduced skewness \(s_3(r_{\parallel},r_{\perp})\) at \(z = 0.1\), shown in real space (top panel) and redshift space (bottom panel). Colors correspond to different \(f\sigma_8\) values. Blue dashed lines indicate iso-volume contours; the diagonal black dashed line marks spherical ellipsoids.}
\label{fig.S3_comparefs8}
\end{figure}

The situation improves when we turn to higher-order statistics. In Figure~\ref{fig.S3_comparefs8}, we show iso-contours of the reduced skewness \(s_3\) in both real and redshift space, colored by the corresponding \(f\sigma_8\) values.

In real space, all models yield nearly identical \(s_3\) maps, making it difficult to distinguish between them. In contrast, in redshift space the contours are clearly separated and follow a consistent trend with increasing \(f\sigma_8\). This enhanced sensitivity highlights the utility of anisotropic skewness as a growth-rate probe.

We also observe a significant difference in the shape of the \(s_3\) map between real and redshift space. The redshift-space suppression of \(s_3\) is strongest at small \(r_{\parallel}\) and \(r_{\perp}\), corresponding to small effective volumes. Moreover, the suppression is anisotropic: for fixed \(r_{\mathrm{eff}}\), oblate ellipsoids (\(r_{\parallel} < r_{\perp}\)) experience greater suppression than prolate ones. This anisotropy reflects the directional nature of FoG distortions and confirms that shape-resolved cumulants offer a richer characterization of redshift-space clustering.

\begin{figure*}
\centering
\includegraphics[width=\textwidth]{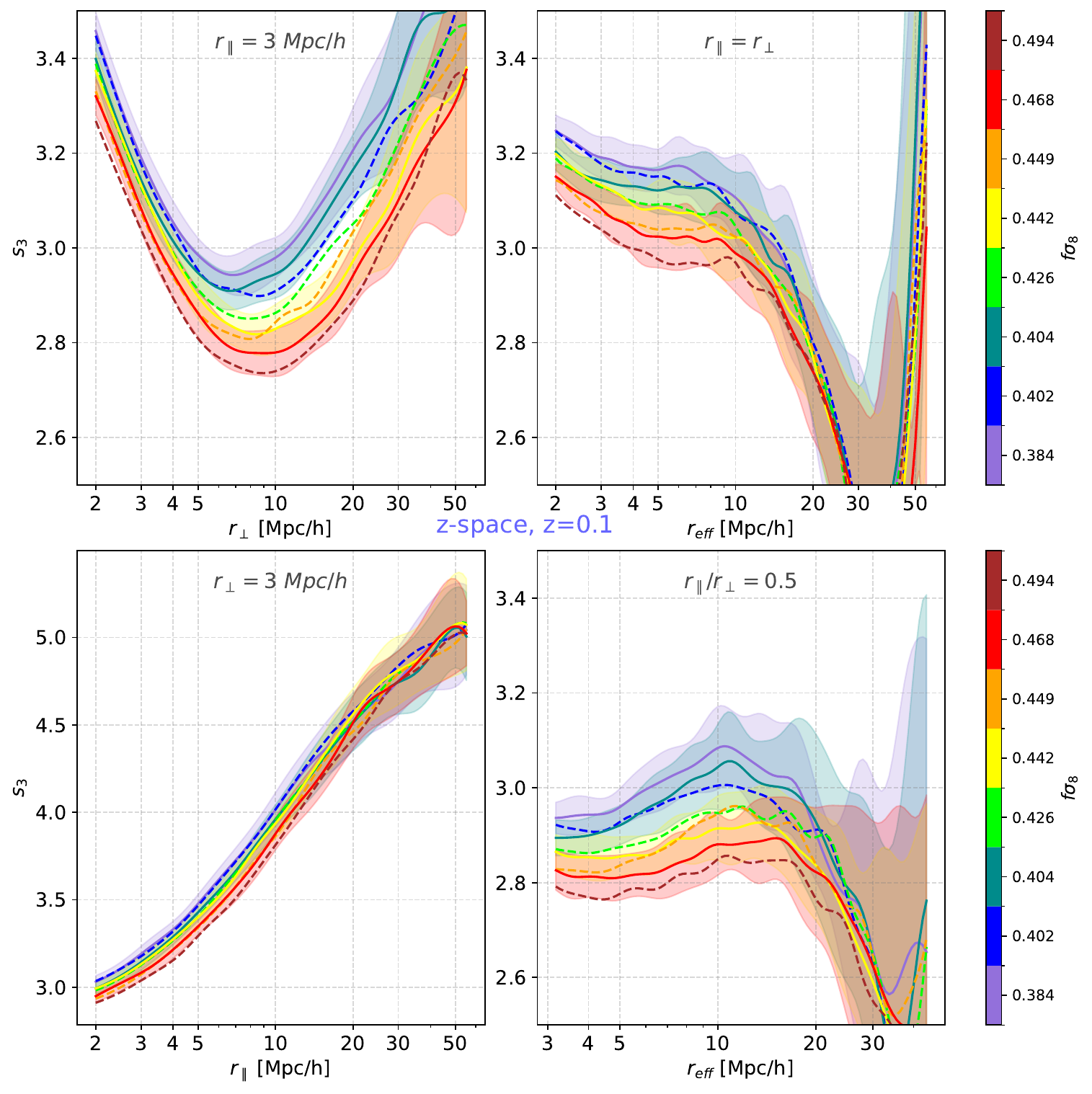}
\caption{Cross-sections of redshift-space skewness \(s_3(r_{\parallel}, r_{\perp})\) at \(z = 0.1\). Panels show: (top left) varying \(r_{\perp}\) at fixed \(r_{\parallel} = 3\,\mathrm{Mpc}/h\); (top right) spherical case \(r_{\parallel} = r_{\perp}\); (bottom left) varying \(r_{\parallel}\) at fixed \(r_{\perp} = 3\,\mathrm{Mpc}/h\); (bottom right) constant axis ratio \(r_{\parallel}/r_{\perp} = 0.5\). 
Solid and dashed lines correspond to different \(f\sigma_8\) values.
For transparency every second model have errors shown (solid lines with shades),
while for the rest we present only values (dashed lines).}
\label{fig.CSec}
\end{figure*}

To explore these trends in more detail, we show cross-sections of \(s_3(r_{\parallel},r_{\perp})\) at various kernel configurations in Figure~\ref{fig.CSec}.
The diversity of shapes across panels underscores the richness of the signal. Differences between models reach the \(1-3\sigma\) level, particularly for cross-sections that probe oblate ellipsoids. For instance, in the bottom right panel, the \(s_3\) signal rises with increasing \(r_{\mathrm{eff}}\) up to \(\sim 10~\mathrm{Mpc}/h\), before falling off at larger scales. This turnover scale is sensitive to \(f\sigma_8\) and is promising for model discrimination.

We conclude that reduced skewness in redshift space, computed using anisotropic ellipsoidal kernels, is a sensitive and robust probe of cosmological growth. In contrast to isotropic VACFs, the directional information retained by ellipsoidal CiC provides access to subtle features of the Cosmic Web and redshift-space distortions, offering a promising avenue for future observational analyses.

\section{Conclusions}\label{sec.concl}

In this work, we introduced a novel generalization of volume-averaged correlation functions (VACFs) and their associated cumulants by extending the standard counts-in-cells methodology to anisotropic, ellipsoidal kernels. These \emph{ellipsoidal volume-averaged correlation functions}, denoted by \(\bar{\xi}_J(r_{\parallel}, r_{\perp})\), and the corresponding reduced cumulants \(s_J(r_{\parallel}, r_{\perp})\), represent a natural configuration-space analogue of the widely used anisotropic two-point function \(\xi(r_p, \pi)\). Our approach distinguishes between directions parallel and perpendicular to the line of sight, allowing for explicit sensitivity to redshift-space distortions (RSD) in higher-order clustering statistics.

We explored the behavior of these estimators in both real and redshift space, using a suite of COLA-based N-body simulations with varying values of \(\Omega_M\) and \(\sigma_8\), and analyzed their ability to capture the imprint of cosmic structure growth and velocity-induced anisotropies. Our main findings can be summarized as follows:

\begin{itemize}
    \item In real space, the VACF \(\bar{\xi}_J(r_{\parallel}, r_{\perp})\) at fixed ellipsoidal volume is maximized for spherical kernels (\(r_{\parallel} = r_{\perp}\)). In redshift space, however, this symmetry is broken by RSD. The preferred kernel shape becomes prolate (\(r_{\parallel} > r_{\perp}\)) on small scales dominated by Fingers-of-God (FoG) effects, and oblate (\(r_{\parallel} < r_{\perp}\)) on large scales where the Kaiser effect enhances LOS compression (Fig.~\ref{fig.W2_contours}).

    \item The shape dependence of higher-order statistics evolves systematically with order \(J\). While lower-order VACFs show strong deviations from iso-volume contours, higher-order \(\bar{\xi}_J\) functions tend to align with iso-volume shapes (Fig.~\ref{fig.WJ_contours_DM}). This suggests that at sufficiently high \(J\), the statistics become effectively volume-dominated and less sensitive to ellipsoidal anisotropy.

    \item In redshift space, variations in \(\sigma_8\) produce a significantly stronger effect on \(\bar{\xi}_2\) than equivalent variations in \(\Omega_M\) (Fig.~\ref{fig.W2_comparefs8}). At small scales, models with higher \(\sigma_8\) show stronger FoG suppression and favor more prolate kernel shapes. At larger scales, the transition to the Kaiser regime occurs at larger effective radii for these models.

    \item While \(\bar{\xi}_J\) does not exhibit a monotonic dependence on the composite growth-rate parameter \(f\sigma_8\), the redshift-space cumulants \(s_J\), particularly the skewness \(s_3\), do. In real space, models with different \(f\sigma_8\) values are nearly indistinguishable, but in redshift space they exhibit clear, coherent differences (Fig.~\ref{fig.S3_comparefs8}).

    \item Cross-sectional analyses of \(s_3(r_{\parallel}, r_{\perp})\) (Fig.~\ref{fig.CSec}) confirm that models with adjacent \(f\sigma_8\) values can be separated at the \(\sim 0.5{-}1\sigma\) level across a range of ellipsoidal configurations, especially in redshift space.
\end{itemize}

Ellipsoidal VACFs and their cumulants offer a powerful extension of the standard counts-in-cells framework, with strong potential for extracting anisotropic clustering information from future redshift surveys. By providing access to directional and morphological dependencies in higher-order statistics, this approach opens a new window onto the physics of structure formation, redshift-space distortions, and the growth of cosmic large-scale structure. Its computational efficiency and general applicability to any cumulant order make it a promising addition to the toolbox of modern cosmological analysis.

Looking ahead, the application of ellipsoidal VACF statistics to galaxy and halo catalogs from simulations and observations represents a particularly promising avenue. Galaxy redshift surveys such as DESI, Euclid, and 4MOST will deliver high-fidelity three-dimensional maps of the cosmic density field, ideally suited to this type of anisotropic analysis. Furthermore, since redshift-space distortions carry complementary information to real-space clustering, our method can serve as an independent probe of the growth of structure—enabling cross-validation of results from standard multipole-based RSD analyses. Finally, the sensitivity of ellipsoidal cumulants to directional anisotropies and nonlinear structure makes them a natural tool for testing extensions of the standard model, including modified gravity theories and alternative dark sector physics. In particular, their ability to access non-Gaussian signatures through higher-order cumulants while resolving geometric distortions makes them ideal for next-generation consistency tests of \(\Lambda\)CDM.

\section{Acknowledgments}
We are  grateful to Enrique Gaztanga for helpful discussions at the early stages of this project. This work is supported by
the Polish National Science Center through grants no.\ 2020/39/B/ST9/03494 and 2020/38/E/ST9/00395.

\bibliography{main}

\end{document}